# Vertical CVD Growth of Highly Uniform Transition Metal Dichalcogenides

Lei Tang[1], Tao Li[2], Yuting Luo[1], Simin Feng[1], Zhengyang Cai[1], Hang Zhang[2], Bilu Liu[1,*], Hui-Ming Cheng[1, 3*]

[1]Shenzhen Geim Graphene Center, Tsinghua-Berkeley Shenzhen Institute and Tsinghua Shenzhen International Graduate School, Tsinghua University, Shenzhen, Guangdong, 518055, P. R. China

[2]Institute of Engineering Thermophysics, Chinese Academy of Sciences, Beijing, 100190, P. R. China

[3]Shenyang National Laboratory for Materials Sciences, Institute of Metal Research, Chinese Academy of Sciences, Shenyang, Liaoning, 110016, P. R. China

Correspondence should be addressed to bilu.liu@sz.tsinghua.edu.cn, hmcheng@sz.tsinghua.edu.cn

**ABSTRACT:** Two-dimensional (2D) transition metal dichalcogenides (TMDCs) have attracted great attention due to their physical and chemical properties that make them promising in electronics and optoelectronics. Because of the difficulties in controlling concentrations of solid precursors and spatially non-uniform growth dynamics, it is challenging to grow wafer-scale 2D TMDCs with good uniformity and reproducibility so far, which significantly hinders their practical use. Here we report a vertical chemical vapor deposition (VCVD) design to grow monolayer TMDCs with a uniform density and high quality over the whole wafer, and with excellent reproducibility. The use of such VCVD design can easily control the three key growth parameters of precursor concentration, gas flow and temperature, which cannot be done in currently widely-used horizontal CVD system. Statistical results show that VCVD-grown monolayer TMDCs including $MoS_2$ and $WS_2$ are of high uniformity and quality on substrates over centimeter size. We also fabricated multiple van der Waals heterostructures by the one-step transfer of VCVD-grown TMDC samples, owning to its good uniformity. This work opens a way to grow 2D materials with high uniformity and reproducibility on the wafer scale, which can be used for the scalable fabrication of 2D materials and their heterostructures.

## INTRODUCTION

2D transition metal dichalcogenides (TMDCs) have attracted significant attention due to their electronic, magnetic, optical and mechanical properties, which have shown potential for use in electronics, optoelectronics, energy conversion and storage, *etc*.[1-4] One major challenge that hinders their practical use is how to produce them in a repeatable and controllable manner with good uniformity and high-quality. Chemical vapor deposition (CVD) has been considered a promising method for preparing 2D TMDCs since it has a good balance between material quality, yield, and cost,[5-7] and up to now has been used to grow monolayer and films like $MoS_2$,[8] $WS_2$,[9] and $WSe_2$.[10] Usually, solid phase precursors like sulfur (selenium) powder and metal oxide powders have been used as precursors to grow TMDCs in a horizontal CVD (HCVD) furnace, which suffers from uncontrollable and non-uniform growth because of uncontrolled precursor concentration, gas flow and temperature. For example, Ling *et al*. have found that the density, lateral size, and morphology of $MoS_2$ grown by the traditional HCVD system were not uniform and depended on the position of the substrate.[11] Van *et al*. have shown that it is difficult to grow TMDCs with good uniformity on centimeter scale substrates even using ultraclean substrates and fresh solid precursors.[12] It is considered that the difficulties in achieving the controllable growth of uniform 2D TMDCs in HCVD systems are mainly caused by the uncontrollable concentrations of solid precursors and spatially non-uniform growth dynamics. In recent years, efforts have been made to solve these issues by choosing alternative precursors and changing the way they are fed into the system. For examples, Kang *et al*. used a metal organic CVD (MOCVD) system with gaseous precursors and grew uniform wafer-scale TMDC films in 24 hours, with domain sizes less than 1 micrometer.[13] By selenizing the solid precursors of a niobium film pre-deposited on the substrates, Lin *et al*. prepared a uniform $NbSe_2$ film with a grain size of tens of nanometers.[14] Nevertheless, it is still a big challenge to grow large domain, uniform, area and high-quality 2D TMDCs.

Here we report the use of vertical CVD (VCVD) to grow monolayer centimeter-scale TMDCs with superb uniformity. The issue of spatially non-uniform growth dynamics is well addressed by using this VCVD system because the vertical design of the furnace and gas flow redistributes the temperature field and ensures the spatial uniformity of the gas flow velocity. In addition, gaseous precursors (*e.g.*, $H_2S$ and an Ar-bubbled metal precursor feed) were used to substitute for the widely-used solid precursors for the growth of TMDCs so as to make the concentration of precursors controllable and steady. As a result of this VCVD design, we achieved the controllable and reproducible growth of 2D TMDCs which could not be achieved using the traditional HCVD system. Statistical results from various characterization techniques show that the VCVD-grown monolayer 2D TMDCs are of high uniformity (including morphology, nucleation density and coverage) and high quality over the centimeter scale. Thanks to the advantages of the VCVD system design, we also fabricated multiple van der Waals heterostructures by the one-step transfer of VCVD-grown large-area 2D TMDCs.

**RESULTS AND DISCUSSION**

The most-widely used horizontal CVD system is named HCVD here to distinguish it from VCVD, since the furnace in HCVD is placed horizontally with a horizontal flow direction. HCVD is widely used to grow various TMDCs in forms of single crystals, films and alloys, *etc*.[15-18] Figure 1a shows a typical HCVD system for growth of 2D TMDCs, where the solid precursors are placed in a horizontal furnace with the substrate downstream. At high temperatures, the precursors are transported and absorbed on the substrate surface for further chemical reactions. The HCVD system typically has a long transport path of the precursor (Figure S1) and there is therefore a gradient of precursor concentration (blue curve in Figure 1c) and a serious drop in the precursor concentration at growth temperature (blue curve in Figure 1d, thermogravimetric analysis (TGA) result in Figure S2), which leads to non-uniform deposition of the TMDCs.[19] A typical result of the growth of a TMDC on a millimeter-scale substrate in a HCVD system is shown in Figure 1e, and is known as position-dependent growth. The as-grown samples show different density distribution of deposited flakes on substrates (Figure S3), similar to previous reports.[11, 12, 20] To overcome these drawbacks, we designed a VCVD system that is mainly composed of an input system, a growth chamber with a specimen holder, an output system and scrubber parts (Figure S4). There are two key features of our VCVD setup (Figure 1b). First, in contrast to the HCVD system, the furnace in the VCVD is vertical with gas flow from top to bottom. The change in the orientation of the furnace may redistribute the temperature and gas field to eliminate position dependence in the growth chamber (red curve in Figure 1c). Second, the VCVD system uses gaseous H₂S and Ar-bubbled metal precursors instead of traditional solid precursors in order to overcome the issue of gradient precursor concentration (red curve in Figure 1d), which could effectively have steady mass flux transportation into the growth chamber. Using this design, we have achieved a uniform growth profile for the VCVD system (Figure 1f).

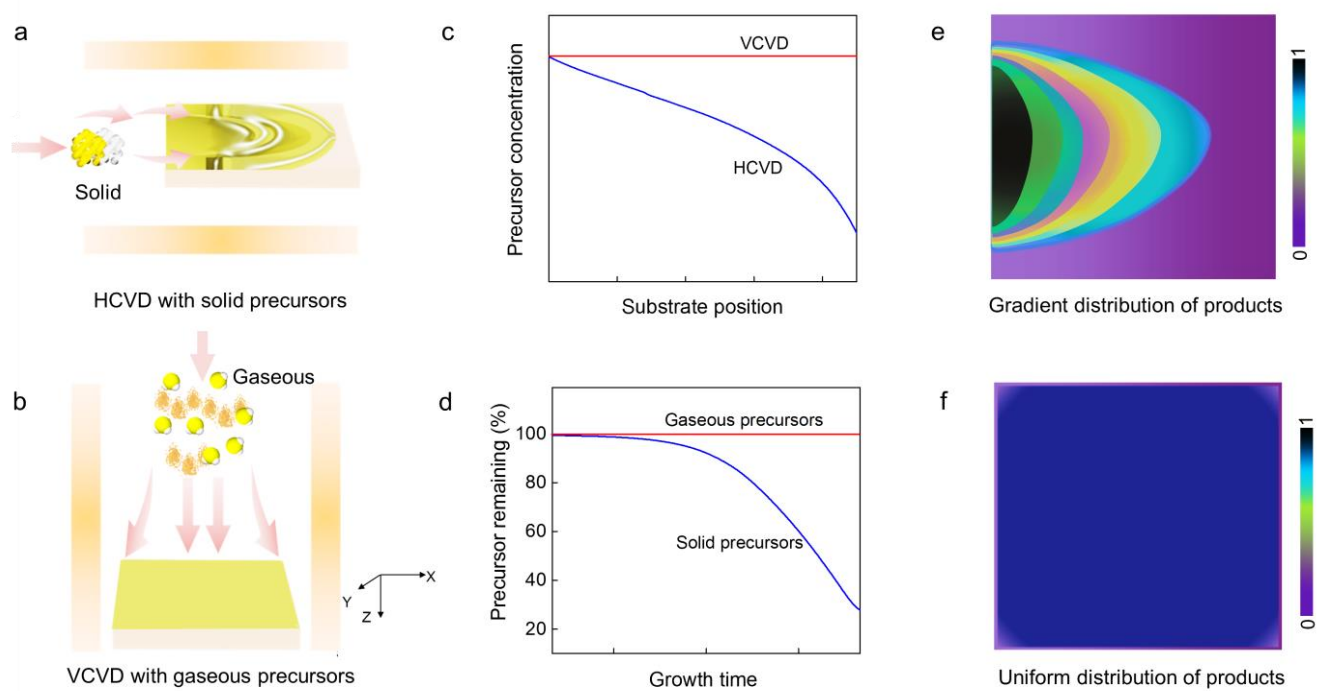

**Figure 1.** Comparison of the key differences between traditional HCVD and VCVD systems. (a, b) Schematics of typical HCVD and VCVD systems using solid precursors and gaseous precursors, respectively. (c, d) Relationships between vapor concentration and substrate position (c), and precursor remaining (mass of precursors) and the growth time (d) in the two different CVD systems. The HCVD system shows substrate position and growth time dependent growth features, while VCVD system does not. (e, f) Growth profiles of the two different CVD systems. HCVD has a gradient diffusion profile with increasing diffusion distance and growth time. In contrast, VCVD exhibits a uniform diffusion profile regardless of diffusion distance and growth time.

We used WS$_2$ as an example to verify the advantages of VCVD design. We tuned the growth parameters in the VCVD to check the distribution of as-grown WS$_2$ domains across the substrate (see details in the Experimental Section). We found that the VCVD-grown WS$_2$ samples have a uniform color compared to the HCVD-grown ones over both sapphire and SiO$_2$/Si substrates with an area of 10×10 mm$^2$ (Figure S5). We characterized the uniformity of the VCVD- and HCVD-grown samples by taking optical microscope images from different positions, which provides macro-views of the sample with acceptable resolution. These optical views (Figure S6) and the real-time Movie (Movie S1) show highly-uniform density and morphology of the VCVD-grown WS$_2$ flakes, while the HCVD-grown sample has a position-dependent deposited distribution (Figure S3). We further carried out the statistical analyses of the uniformity of VCVD samples, including nucleation density, coverage and lateral size. Figure 2a and 2b show the corresponding results of nucleation density and coverage of the as-grown samples, which show uniform distribution. The average domain size of the as-grown TMDCs is 9.7 μm (Figure S7), which is larger than the previous MOCVD results.[13, 21, 22] Besides WS$_2$, we also synthesized uniform MoS$_2$ samples (Figure S8). We performed growth experiments 100 times and 90% of the experiments have a similar density of TMDC growth triangles, showing good reproducibility. Overall, these results show the effectiveness of the VCVD method in producing spatially uniform 2D TMDCs.

We then focus on the quality and uniformity of the VCVD-grown TMDC domains. Figures 2c and 2d show the statistical Raman of the spacing between the E$^1_{2g}$ and A$_{1g}$ Raman peaks and photoluminescence (PL) intensity of WS$_2$ domains on the 10×10 mm$^2$ sapphire substrate, the results show a constant value for all of them. Figure 2e-2g show the uniform Raman intensity maps of the E$^1_{2g}$, A$_{1g}$ peaks and PL bandgap, which are located at 352 cm$^{-1}$ and 419 cm$^{-1}$ for monolayer WS$_2$ (Figure

S9a) and 1.98 eV (Figure S9b). It is worth noting that the intensity maps are uniform across all domains, indicating uniformity in the number of layers and crystalline quality of the sample. Furthermore, all of the UV visible absorption spectra show an absorption peak at 1.98 eV at seven different positions on the samples (Figure S10), suggesting high crystalline quality and uniformity.[23] XPS was used to measure the atomic ratio of sulphur/tungsten (S/W) as 1.99 (Figure S11), which shows good stoichiometry and quality. Meanwhile, for VCVD-grown $MoS_2$, typical Raman spectra show the positions of $E^1_{2g}$ and $A_{1g}$ peaks located at 380 cm$^{-1}$ and 403 cm$^{-1}$ (Figure S12a), and the corresponding PL peak at 1.83 eV, indicating monolayer $MoS_2$ flakes have been grown (Figure S12b). Overall, the above spectroscopy results confirm the high uniformity and crystalline quality of the VCVD-grown TMDC domains.

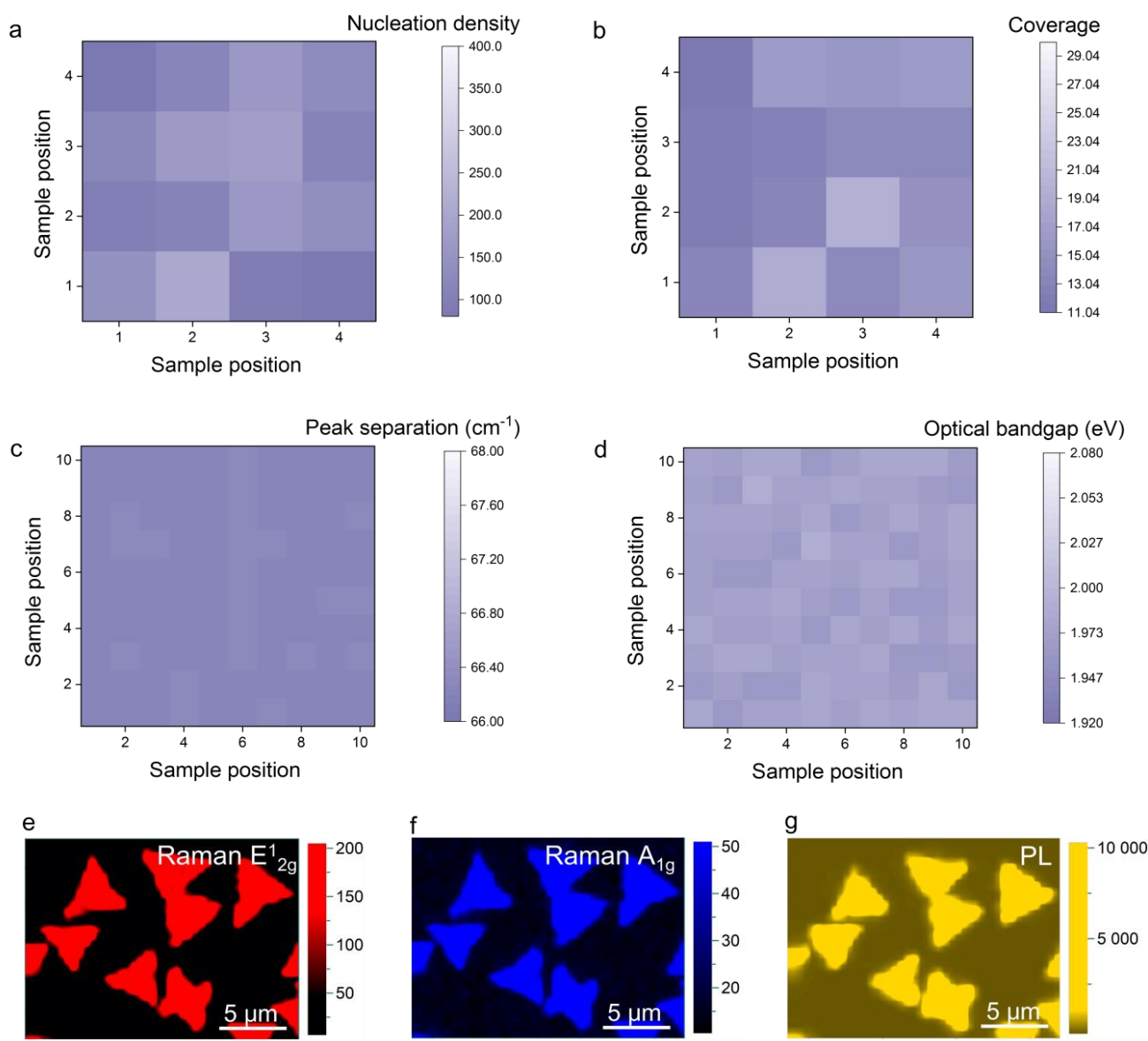

**Figure 2.** Optical microscope and spectroscopic characterization of the VCVD-grown monolayer WS$_2$ domains with high uniformity and quality. (a, b) Statistics of the nucleation density and coverage of WS$_2$ domains at different 16 positions according to the optical images taken on a 10×10 mm$^2$ sapphire substrate. (c) The statistical peak separation of the E$_{2g}^1$ peak and A$_{1g}$ peak from the respective Raman spectrum, which was obtained from 100 different positions. (d) The statistical results of PL peak position plotted as a function of 100 different positions. (e-g) The corresponding Raman intensity maps of the E$_{2g}^1$ and A$_{1g}$ peaks and PL intensity map at ~626 nm.

We analyzed how gas flow and temperature were affected by the orientation of the furnace in VCVD and HCVD systems. It is known that gas flow dynamics[14, 24-26] and temperature distribution[6, 27] play important roles in the CVD growth of 2D materials, including uniformity, grain size, thickness, growth rate, structure and morphology. To clarify the importance of these factors for producing TMDCs with a high uniformity, we performed a computational fluid dynamics (CFD) simulation to study the distribution of gas flow velocity and temperature in VCVD and HCVD systems.[28] The gas flow in the HCVD chamber is turbulent and non-uniform which increases the chance of forming random nucleation sites and results in unsteady growth conditions (Figure 3c). In contrast, in the VCVD system the velocity of gas flow is uniform and remains steady throughout the growth chamber (Figure 3a). In addition, the temperature distributions in these two systems show similar results (Figures 3b and 3d) in accordance with the distributions of gas flow. We further considered the growth substrate is perpendicular to the direction of gas flow in the HCVD (Figure S13), which indicates the formation of vortexing gas of precursor concentration along the substrate. And the SEM image of the vertically-standing TMDC nanosheets grown on $SiO_2$/Si substrate is shown in Figure S14. After comparing all the designs, the importance and advantages of VCVD design are further proved. Basically, the CVD process involves typical heterogeneous nucleation reactions.[29] Heterogeneous nucleation forms at preferential sites such as phase boundaries, surfaces of the container, substrates, or impurities like dust and particles. At such sites, the effective surface energy is low, which lowers the free energy barrier and facilitates nucleation for TMDC growth and evolution. The free energy needed for heterogeneous nucleation is equal to the product of homogeneous nucleation and a function of the contact angle ($\theta$, Figure S15),

$$\Delta G^*_{heterogeneous} = \Delta G^*_{homogeneous} * f(\theta) \quad (1)$$

By analyzing the process of vapor phase deposition, we have obtained the following formula,

$$\Delta G^*_{heterogeneous} = \Delta G^*_{homogeneous} * f(\theta) \sim \frac{\gamma^3_{gas-sol}}{\Delta G^2_V} * f(\theta) \quad (2)$$

where $\gamma$ is the critical radius of the interface between the gas and the solid substrate, and $\Delta G^2_V$ is the total free energy and the indicator of gas pressure. When the flow rate is high, $\Delta G^2_V$ is increased, which leads to a decrease of $\Delta G^*_{heterogeneous}$, thus causing a higher nucleation density. In a HCVD system, the gas flow of the solid vapor is not uniform in the furnace, and the flow rate at the center is higher than at the edge (Figure S16), which produces a lower $\Delta G^*_{heterogeneous}$ and a higher nucleation density at the center. In contrast, the flow rate in the growth chamber is much steadier and more uniform in the VCVD system, which results in a constant value of $\Delta G^*_{heterogeneous}$ and a constant possibility of nucleation density. Therefore, VCVD systems with gaseous precursors show clear advantages for the growth of uniform TMDCs.

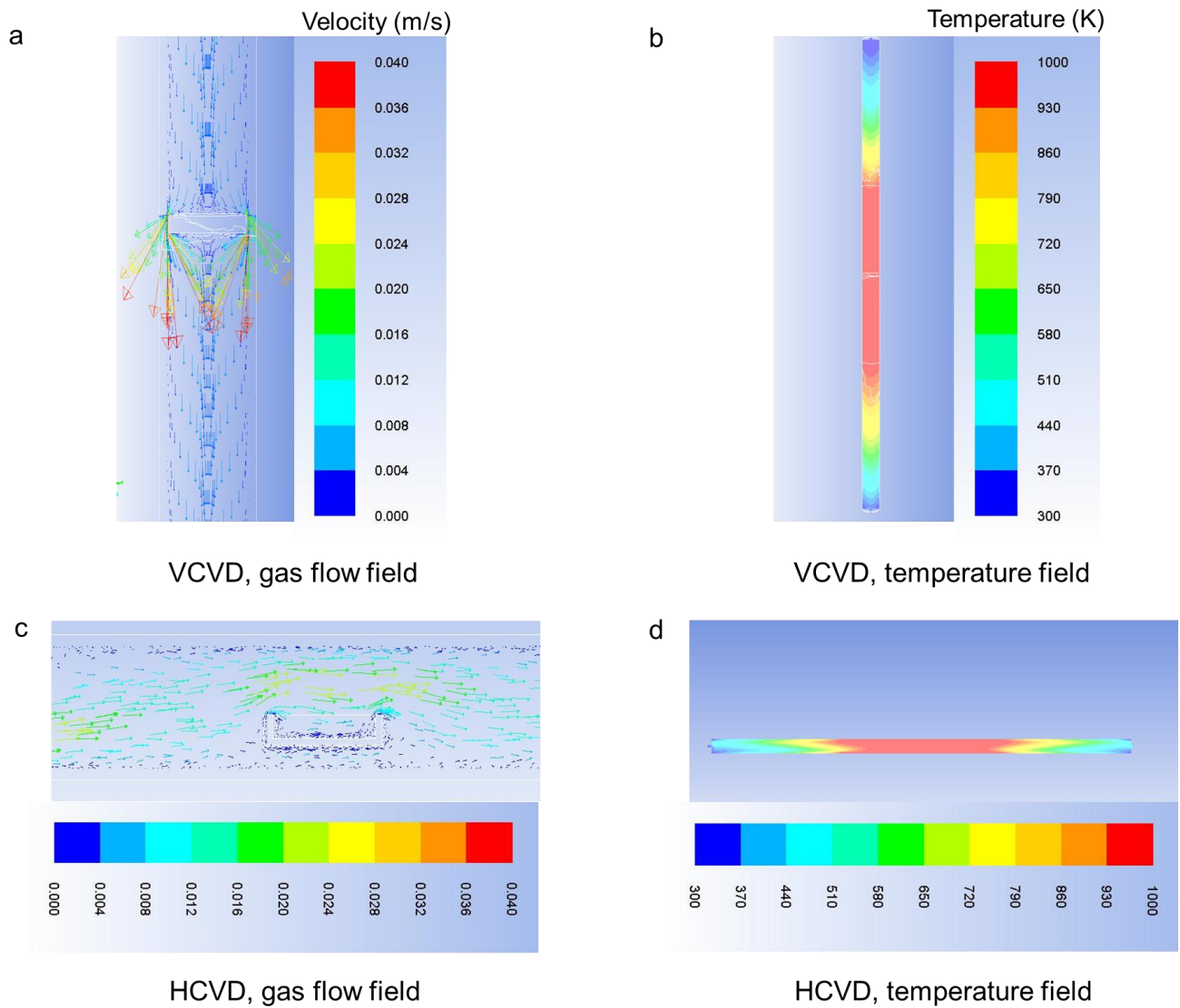

**Figure 3.** CFD simulations of the distribution of gas flow velocity and temperature in VCVD and HCVD systems. (a, c) Simulation of the gas flow velocity distribution in VCVD (a) and HCVD (c). (b, d) Simulation of the temperature distribution in VCVD (b) and HCVD (d).

The availability of VCVD-grown large-area uniform 2D TMDCs provides a platform for the fabrication of multiple TMDC heterostructures by a one-step transfer process. Figure 4a shows a schematic of the thermal release tape (TRT) transfer of as-grown TMDCs (see details in the Experimental Section). We have fabricated multiple van der Waals heterostructures including

WS$_2$/MoS$_2$, WS$_2$/graphene and MoS$_2$/graphene on arbitrary substrates (Figure 4b-4f). The corresponding Raman and PL spectra of the heterostructures are shown in Figure 4g-4l. They show typical Raman peaks of monolayer WS$_2$, MoS$_2$ and few-layer graphene in the heterostructures, such as the E$_{2g}^1$ and A$_{1g}$ peaks of WS$_2$, the E$_{2g}^1$ and A$_{1g}$ peaks of MoS$_2$, and the G and 2D bands of graphene (Figure 5g-5i). The PL spectra acquired from the heterostructures show peaks at wavelengths of 630 nm (1.96 eV direct exciton transition energy of monolayer WS$_2$) and 680 nm (1.82 eV direct exciton transition energy of monolayer MoS$_2$). These PL spectra indicate that the presence of graphene does not strongly quench the PL intensity of the TMDCs (Figures 4k and 4l). In addition, we do not observe a strong direct exciton peak at 875 nm (1.42 eV) in the MoS$_2$/WS$_2$ heterostructure (Figure 4j). These results show the successful fabrication of various heterostructures in a simple one-step transfer process, but not the same as those in the cases of TMDCs directly grown on graphene[30] or one-step CVD-grown TMDC vertical heterostructures.[6]

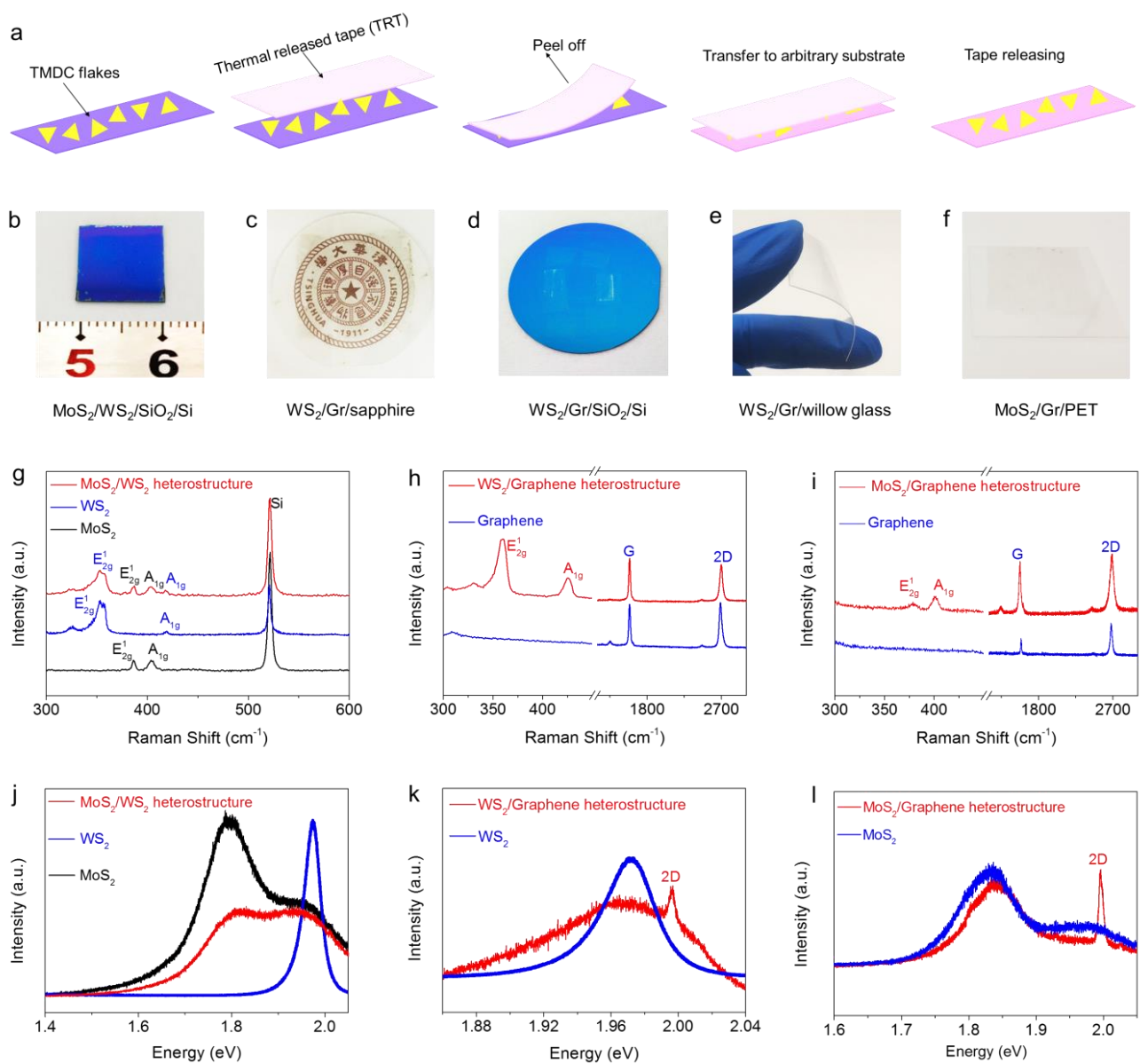

**Figure 4**. TRT transfer for fabricating multiple van der Waals heterostructures from VCVD grown TMDCs. (a) Schematic of the TRT transfer process. (b-f) Photographs of MoS$_2$/WS$_2$, WS$_2$/graphene and MoS$_2$/graphene van der Waals heterostructures on different substrates. 'Gr' represents graphene. (g-i) Raman spectra of the MoS$_2$/WS$_2$, WS$_2$/graphene and MoS$_2$/graphene heterostructures. (j-l) PL spectra of the MoS$_2$/WS$_2$, WS$_2$/graphene and MoS$_2$/graphene heterostructures.

**CONCLUSION**

We have designed a VCVD system for the growth of TMDCs with wafer-scale uniformity and excellent reproducibility. The VCVD system substantially addresses the challenging issues of spatially non-uniform growth dynamics and uncontrollable solid precursors in widely-used HCVD systems, and thus ensures 2D TMDCs with high uniformity, quality and reproducibility. Moreover, by combining the heterogeneous nucleation mechanism and the CFD simulations, we have shed new light on the mechanism and the effect of the key physical parameters like precursor concentration, gas flow, temperature and free energy of heterogeneous nucleation, which dominate the growth behavior of 2D TMDCs in the CVD process. Using the VCVD-grown highly-uniform 2D TMDCs, we fabricated different TMDC heterostructures by a one-step transfer process. Our work opens a powerful way to grow large-scale uniform TMDCs and various TMDC heterostructures.

**EXPERIMENTAL SECTION**

*Materials and chemicals.* Sulphur powder (99.95%, Alfa Aesar, USA), tungsten (VI) chloride ($WCl_6$, 99.9%，Shanghai Macklin Biochemical Co., Ltd., China), molybdenum (VI) chloride ($MoCl_6$, 99.9%，Shanghai Macklin Biochemical Co., Ltd., China), molybdenum (VI) oxide powder ($MoO_3$, 99.99%, Alfa Aesar, USA), tungsten (VI) oxide powder ($WO_3$, 99.99%, Alfa Aesar, USA), PDMS tape (200 μm thickness, Hangzhou Bald Advanced Materials Co., Ltd., China), thermal release tape (Nanjing MKNANO Tech. Co., Ltd., China), c-plane sapphire substrate (Nanjing MKNANO Tech. Co., Ltd., China), $SiO_2$/Si substrate (300 nm oxide layer thickness, Hefei Kejing Materials Technology Co., Ltd., China), isopropyl alcohol (Analytical Reagent, Shanghai Macklin Biochemical Co., Ltd., China) were used as-received.

***Growth of 2D TMDCs in a VCVD system.*** In our experiments, growth was conducted in a homemade atmospheric pressure vapor deposition vertical furnace with a 1.5-inch-diameter quartz tube (OTF-1200X, Hefei Kejing Materials Technology Co., Ltd., China). $H_2S$ gas was used as the sulphur source. A metal-containing precursor ($WCl_6$ or $MoCl_6$) was dissolved in isopropyl alcohol and bubbled into the growth chamber by Ar (99.99%) as the metal source. The growth substrate (sapphire or $SiO_2$/Si) was placed at the center of the furnace with a specimen holder. The furnace was heated to the growth temperature of 850-950 °C with a heating rate of 30 °C $min^{-1}$ and kept there for 30-90 min for $WS_2$ growth, followed by cooling. During heating and cooling, Ar was used (800 sccm, standard cubic centimeters per minute). During growth, Ar (800 sccm), $H_2S$ (20 sccm), $H_2$ (20 sccm) and $WCl_6$ (bubbled by 20 sccm of Ar) were introduced for the growth of $WS_2$. The $MoS_2$ was grown using the same method but using $MoCl_6$ as the molybdenum source.

***Growth of 2D TMDCs in a HCVD system.*** As a comparison experiment, growth was conducted in a homemade two-zone HCVD furnace with a 1-inch diameter quartz tube (TF55035C-1, Lindberg/Blue M, Thermo Fisher Scientific, USA). Sulfur powder (100 mg) was located in the first upstream zone, and $WO_3$ (3-10 mg) with a facedown $SiO_2$/Si substrate (or a c-plane sapphire substrate) was placed downstream (8-10 cm) for $WS_2$ growth. First, the tube was flushed with Ar for 30 min. The furnace was then heated to 780 -950 °C with a heating rate of 40 °C $min^{-1}$ where is was kept for 1-5 min for $WS_2$ growth. After growth, the furnace was cooled to room temperature while 50 sccm of Ar was introduced during the whole process. The $MoS_2$ was grown using the same method but $MoO_3$ was used to replace the $WO_3$ precursor.

***Fabrication of the heterostructures.*** First, thermal release tape was attached to the TMDC/sapphire and pressed for 60 min. The tape/TMDC was peeled off the tape/TMDC/sapphire carefully and slowly using a micromechanical stage. After that, the tape/TMDCs was attached to an arbitrary substrate covered with a graphene film. After heating in air at 140 °C for 90-240 s, the tape was lifted off from the heterostructures.

***Characterization of as-grown 2D TMDCs.*** An optical microscope (Carl Zeiss Microscopy, Germany) and an SEM (Hitachi SU8010, Japan) were used to observe the morphology and uniform distribution of the TMDCs. Raman and PL spectroscopy were performed under 532 nm laser excitation (Horiba LabRAM HR Evolution, Japan). Structural and chemical analyses of the samples were performed by XPS (Thermo Scientific K-Alpha XPS, using Al (Kα) radiation as a probe, USA) and UV-Vis-NIR absorption (Perkin-Elmer Lambda 950 spectrophotometer, USA). Thermogravimetric analysis (Mettler Toledo TGA2, USA) was used to check the remaining mass of metal oxide precursors ($MoO_3$) with increasing temperature and time.

## ASSOCIATED CONTENT

**Supporting Information**

The Supporting Information is available free of charge on the ACS Publications website.

## AUTHOR INFORMATION

**Corresponding Authors**

[*]bilu.liu@sz.tsinghua.edu.cn

[*]hmcheng@sz.tsinghua.edu.cn

**ORCID**

Bilu Liu: 0000-0002-7274-5752

Hui-Ming Cheng: 0000-0002-5387-4241

**Notes**

The authors declare no competing financial interest.

**ACKNOWLEDGMENTS**

We thank Mingqiang Liu and Dr. Usman Khan for helpful discussions, and Hao Xu for drawing the schematics. We acknowledge the supports by the National Natural Science Foundation of China (Nos. 51722206, 51920105002, 51991340, and 51991343), the National Key R&D Program (2018YFA0307200), Guangdong Innovative and Entrepreneurial Research Team Program (No. 2017ZT07C341), the Bureau of Industry and Information Technology of Shenzhen for the "2017 Graphene Manufacturing Innovation Center Project" (No. 201901171523), and the Development and Reform Commission of Shenzhen Municipality for the development of the "Low-Dimensional Materials and Devices" discipline.